\documentclass[sigplan, screen]{acmart}

\renewcommand\footnotetextcopyrightpermission[1]{} 
\usepackage{cleveref}
\usepackage{siunitx}            
\usepackage{courier}            
\usepackage{
comment}
\usepackage[scaled]{helvet} 
\usepackage{url}                  
\usepackage{listings}
\usepackage{enumitem}      
\usepackage{xspace}

\begin{document}

\title{bpftime: userspace eBPF Runtime for Uprobe, Syscall and Kernel-User Interactions}
%

\author{Yusheng Zheng}
\affiliation{%
  \institution{eunomia-bpf Community}
  \city{Shanghai}
  \country{China} 
}
\email{yunwei356@gmail.com}

\author{Tong Yu}
\affiliation{%
  \institution{eunomia-bpf Community}
  \city{Wuhan}
  \country{China} 
}
\email{yt.xyxx@gmail.com}

\author{Yiwei Yang}
\affiliation{%
  \institution{University of California, Santa Cruz}
  \streetaddress{Computer Science Engineering}
  \city{Santa Cruz}
  \state{California}
  \country{USA} 
  \postcode{95060}
}
\email{victoryang00@ucsc.edu}

\author{Yanpeng Hu}
\affiliation{%
  \institution{ShanghaiTech University}
  \streetaddress{Computer Science Department}
  \city{Shanghai}
  \state{Shanghai}
  \country{China} 
  \postcode{20439}
}
\email{huyp@shanghaitech.edu.cn}

\author{Xiaozheng Lai}
\affiliation{%
  \institution{South China university of technology}
  \streetaddress{computer  science & engineering course}
  \city{Guangzhou}
  \state{Guangdong}
  \country{China} 
}
\email{laixz@scut.edu.cn}

\author{Andrew Quinn}
\affiliation{%
  \institution{University of California, Santa Cruz}
  \streetaddress{Computer Science Engineering}
  \city{Santa Cruz}
  \state{California}
  \country{USA}
  \postcode{95060}
}
\email{aquinn1@ucsc.edu}

\begin{abstract}
In kernel-centric operations, the uprobe component of eBPF frequently encounters performance bottlenecks, largely attributed to the overheads borne by context switches. Transitioning eBPF operations to user space bypasses these hindrances, thereby optimizing performance. This also enhances configurability and obviates the necessity for root access or privileges for kernel eBPF, subsequently minimizing the kernel attack surface.

This paper introduces bpftime, a novel user-space eBPF runtime, which leverages binary rewriting to implement uprobe and syscall hook capabilities. Through bpftime, user-space uprobes achieve a 10x speed enhancement compared to their kernel counterparts without requiring dual context switches. Additionally, this runtime facilitates the programmatic hooking of syscalls within a process, both safely and efficiently. Bpftime can be seamlessly attached to any running process, eliminating the need for either a restart or manual recompilation. Our implementation also extends to interprocess eBPF Maps within shared memory, catering to summary aggregation or control plane communication requirements. Compatibility with existing eBPF toolchains such as clang and libbpf is maintained, not only simplifying the development of user-space eBPF without necessitating any modifications but also supporting CO-RE through BTF. Through bpftime, we not only enhance uprobe performance but also extend the versatility and user-friendliness of eBPF runtime in user space, paving the way for more efficient and secure kernel operations. 
\end{abstract}

\maketitle

\section{Introduction}
The Extended Berkeley Packet Filter (eBPF) framework has evolved as a fundamental element in modern kernel programming, providing a streamlined approach to execute user-defined programs within the kernel's privileged domain. Initially designed for networking tasks, eBPF has expanded its horizons to include various kernel subsystems, facilitating performance monitoring, security enforcement, and more. Operating within the kernel, eBPF enables quick observability and data collection. However, it comes with certain performance and security drawbacks. Notably, the context switch overheads associated with Uprobe components hinder performance, while the required privilege access could broaden the kernel attack surface, potentially leading to container escapes\cite{he2023cross} or kernel exploits\cite{lim2023unleashing}. These challenges prompted the exploration of userspace eBPF runtimes, aiming to leverage the advantages of eBPF while addressing its downsides.

The transition towards userspace eBPF began with projects like uBPF\cite{ubpf} and rbpf\cite{rbpf}. Although groundbreaking, these projects faced hurdles regarding performance, flexibility, and feature support. For instance, these early userspace runtimes lacked essential attach facilities like uprobes and syscalls, required manual compilation and program restarts for integration, and had an insufficient hashmap implementation. They also necessitated specific toolchains for program development, thereby posing barriers to wider adoption and showing limited kernel compatibility. Despite these shortcomings, they demonstrated the potential and emphasized the need for a more robust, performance-optimized, and feature-rich userspace eBPF runtime. The employment of userspace eBPF in significant projects further highlighted its benefits and the urgency for refined userspace eBPF runtimes. For example, DPDK utilized userspace eBPF to improve flexibility and performance in network processing\cite{dpdk-ebpf}, while the ongoing eBPF for Windows\cite{windows-ebpf} project displayed the cross-platform potential of eBPF, suggesting a substantial extension in applications from network processing to blockchain smart contract execution. This range of applications, along with the limitations of existing userspace runtimes, paved the way for a more advanced solution. Papers like RapidPatch\cite{he2022rapidpatch} and Femto-Containers\cite{zandberg2022femto} unveiled frameworks for patch propagation and secure deployment on embedded and IoT devices respectively, further displaying eBPF's versatile applications. These practical projects and scholarly discussions collectively highlight the urgent need for a sophisticated userspace eBPF runtime like bpftime, aimed at surmounting the identified challenges and fostering further integration and adoption of userspace eBPF across varied applications.

Introducing bpftime—a cutting-edge userspace eBPF runtime developed on the foundation of LLVM's JIT/AOT technology, addressing the earlier identified challenges while significantly enhancing the performance of Uprobe and Syscall hooks. Unlike its forerunners, bpftime not only boosts Uprobe execution speed to more than 10x but also pioneers in providing programmatic Syscall hooking, shared memory maps, and a smooth interface with popular eBPF toolchains such as libbpf and clang. Remarkably, this runtime can be incorporated into any ongoing process without requiring a restart or manual recompilation, showcasing the much-needed flexibility and ease of integration.

In summary, the key contributions from our work include:







\begin{enumerate}
    \item \textbf{High-Performance, general-purpose Userspace eBPF Runtime compatibility with existing eBPF echosystem}: 
    We unveil \textit{bpftime}, a fresh userspace eBPF runtime, with a fast LLVM JIT and share-memory maps. Our endeavor retains compatibility with existing eBPF toolchains and libraries like clang and libbpf, and can run existing eBPF applications without modification.

    \item \textbf{Programmatic attaching mechanism and Seamless Runtime Injection for userspace eBPF}: 
    Bpftime enables proficient programmatic syscall hooking and uprobe attaching within a process, compatible with kernel eBPF. Additionally, it supports seamless runtime injection into any running process, eliminating the need for cumbersome restarts or manual recompilation. Through an innovative binary rewriting technique, it achieves a substantial more than 10x reduction in Uprobe overhead when compared to kernel uprobes.

    \item \textbf{Working together with kernel eBPF and Extensible Feature Set}: Enable seamless load userspace eBPF from kernel, and using kernel eBPF maps to cooperate with kernel eBPF programs. It can also support new eBPF types from kernel eBPF, especially the  FUSE BPF filter types for better supporting the userspace filesystem helper function.
\end{enumerate}



\section{Background}

In this section, we will delve into the foundational concepts that underpin our discussion in the subsequent sections.
\subsection{eBPF in Kernel}

The Extended Berkeley Packet Filter (eBPF) has blossomed into a multifaceted technology in kernel programming. Evolving from the Berkeley Packet Filter (BPF), eBPF has broadened its horizon from mere packet filtering to a plethora of kernel functionalities, offering a high-performance conduit for user-defined programs to interact within the kernel's privileged domain. Initially tailored for networking tasks, eBPF has now ventured into various kernel subsystems, enabling applications in performance monitoring, security enforcement, and beyond. eBPF operates by allowing pre-compiled programs to be executed within an event-driven framework in response to kernel and userspace events, all under the vigilant scrutiny of a verifier ensuring the kernel's stability and security.

\begin{figure}[t]
\centering
\includegraphics[width=0.5\textwidth]{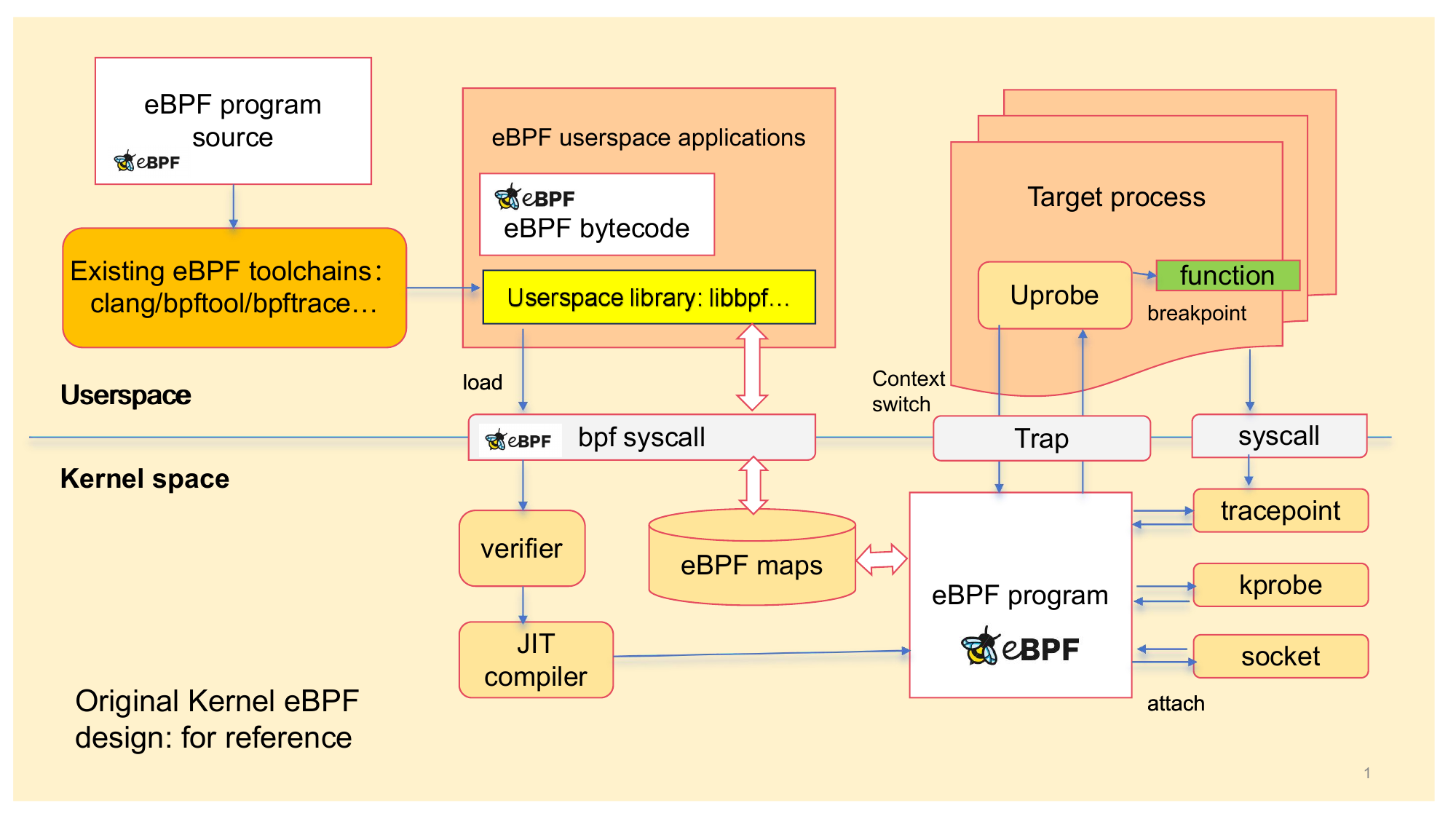}
\caption{The Workflow of kernel eBPF runtime}\label{fig:workflow}
\end{figure}

A typical kernel eBPF runtime is depicted in \cref{fig:workflow}. The eBPF program source is compiled using toolchains like clang, and loaded from a userspace eBPF application to the kernel via the BPF syscall, aided by libraries such as libbpf. Post verification, the BPF Just-In-Time (JIT) compiler transmutes the BPF bytecode to machine code for execution. The userspace eBPF application also orchestrates maps to steer the kernel eBPF behavior, aggregate or encapsulate the kernel BPF data, and relay information from the kernel to the user space. Moreover, it can tether BPF programs to events like kprobe, socket, syscall tracepoints, uprobe, and more, furnishing a robust mechanism for real-time interaction and data retrieval.

However, the kernel environment is not without its caveats. Security concerns are paramount -- eBPF programs operate in kernel mode, necessitating root access, which enlarges the attack surface, thereby brewing risks such as container escapes. Inherent vulnerabilities within eBPF could potentially be harnessed for kernel exploits. Furthermore, the verifier curtails the operations of eBPF, and any configuration alteration or enhance demands kernel changes\cite{Zhong22}. Although the community has pondered over unprivileged eBPF, the features it offers in this guise are severely curtailed, lacking the ability to trace syscalls and userspace functions, thus diminishing its appeal.

\subsection{Uprobe and Syscall Tracepoints}

Uprobes and syscall tracepoints are fundamental mechanisms that enable interaction and introspection within the kernel, particularly when used alongside eBPF. 

\textbf{Uprobes}, which stand for user-level dynamic tracing \cite{keniston2007ptrace}, and are part of perf-tools, allow for dynamic instrumentation of any routine in the user space, facilitating the collection of crucial data for performance analysis and fault diagnosis. Unlike kprobes, which operate within the kernel, uprobes trap into the kernel using the int3 instruction and execute in the kernel space with kernel eBPF, incurring a significant overhead due to the two context switches required - one into the kernel and another back out to the user space. This overhead can often result in a 10 to 20 times slowdown, make it not suitable for real-time monitoring in latency-sensitive applications. Despite the overhead, the combination of uprobes and the eBPF framework provides a powerful tool for non-intrusive tracing of user space functions without necessitating recompilation or restarting user space processes. This has led to its wide adoption in production environments. For instance, uprobes have been utilized for SSL/TLS data capture \cite{ecapture}, memory allocation monitoring and goroutine tracing\cite{bcc} in notable open-source projects. Additionally, they play a significant role in Distributed Tracing Troubleshooting in Zero Code environments, aiding in the identification and resolution of system issues with minimal intrusion into running processes\cite{shen2023network}.

\textbf{Syscall tracepoints} provide a simple way to intercept and modify system calls throughout the system, becoming powerful tools for security and monitoring when used with eBPF. With eBPF, syscall tracepoints can be set up to react to or change system call parameters and return values, allowing detailed control and monitoring of system interactions. In action, when a system call is made, the relevant syscall tracepoint is triggered, starting the connected eBPF program. This eBPF program can inspect the system call parameters, make decisions based on its rules, and even change the return value of the system call. For instance, an eBPF program can be set to block certain system calls by changing the return value to an error code, offering a straightforward way for system call filtering crucial for security. Tracepoints were among the early features brought in with eBPF and have gained wide usage due to their reliability. Unlike kprobes, syscall tracepoints provide a more stable interface and can be used reliably across different kernel versions, making them a favored choice for many. The clear-cut interface of syscall tracepoints reduces the chance of inconsistencies across different kernel versions, establishing the pairing of syscall tracepoints and eBPF as a strong solution for system-wide interception and alteration of system calls.

\subsection{Userspace Virtual Machines: eBPF and Wasm}

The realm of userspace virtual machines has seen the advent of two notable runtimes: eBPF (Extended Berkeley Packet Filter) and Wasm (WebAssembly). Each of these runtimes has distinct design philosophies and operational principles, which cater to varying use-cases and operational requirements.

eBPF, originally rooted in the kernel space, migrated to userspace to alleviate some challenges associated with kernel-based execution. The migration was pioneered by projects like uBPF and rbpf, which aimed to harness the robust capabilities of eBPF while ensuring safer and more convenient user-space operations \cite{ubpf, rbpf}. However, these initial attempts encountered hurdles including the lack of comprehensive support for uprobes and syscalls, the need for manual compilation and program restarts for integration, and an underwhelming hashmap implementation. Despite these limitations, these projects showcased the potential of userspace eBPF runtimes and paved the way for further advancements. For instance, frameworks like RapidPatch and Femto-Containers leveraged eBPF for patch propagation and secure deployment on embedded and IoT devices, demonstrating the versatile applications of eBPF \cite{rapidpatch, femto}. Moreover, projects like DPDK and the ongoing eBPF for Windows initiative indicate substantial progress towards utilizing eBPF benefits across a broader spectrum of applications and platforms \cite{dpdk-ebpf, windows-ebpf}. eBPF prioritizes performance, employing a verifier to ensure security, albeit placing it as a secondary concern.

On the other hand, Wasm is a well-established userspace virtual machine, extensively adopted as a plugin system and a lightweight containerization solution. Its wide language support enables the execution of entire libraries or processes, fostering a rich ecosystem for development. However, Wasm presents some limitations, particularly when it comes to integration. Manual integration is often required, making it less adaptable to API version changes. It relies heavily on underlying libraries for complex operations, such as Wasi-nn, and external APIs like Wasi necessitate additional validation and runtime checks, leading to high performance costs. Unlike eBPF, Wasm emphasizes security, employing Software Fault Isolation (SFI) for security, even at the expense of some runtime overheads.

Both eBPF and Wasm exemplify the strides made in userspace virtual machine technologies, each with its unique strengths and challenges. Their comparative analysis provides a richer understanding of the landscape and the trade-offs involved in leveraging these runtimes for different application domains.
\section{Design}

\subsection{Goals}

The design of \textit{bpftime} is driven by the goal to harness the capabilities of eBPF efficiently while addressing the limitations associated with its traditional in-kernel execution. The primary objectives guiding the development of \textit{bpftime} are outlined as follows:

\begin{enumerate}
    \item \textbf{Userspace Execution of eBPF:} Transitioning eBPF execution to the user space is central to minimizing context switch overhead and offering a more configurable environment. By moving eBPF out of the kernel and into the user space, \textit{bpftime} seeks to alleviate the performance bottlenecks while retaining the powerful features of eBPF.
    
    \item \textbf{Kernel Compatibility:} Ensuring compatibility with the existing kernel interfaces is crucial for \textit{bpftime} to support the current toolchains and applications built for eBPF. This compatibility facilitates a smooth transition for developers and systems already leveraging eBPF, without necessitating a steep learning curve or major modifications to existing codebases.
    
    \item \textbf{Support for Uprobe and Syscall Hooks:} Extending support for uprobes and syscall hooks is fundamental to covering a broad spectrum of eBPF use cases. \textit{bpftime} aims to provide a robust framework where these hooks can be dynamically attached to target processes without the need for restarts or recompilations, thereby enhancing the flexibility and ease of use for dynamic tracing and monitoring tasks.
    
    \item \textbf{Performance Optimization and Platform Extensibility:} Achieving high performance is a cardinal goal for \textit{bpftime}, alongside ensuring its extensibility across multiple platforms. By optimizing the runtime for speed and expanding platform support, \textit{bpftime} aspires to cater to a wider audience and a variety of use cases, embodying a versatile and high-performance userspace eBPF runtime.
\end{enumerate}

The aforementioned goals frame the foundational design principles of \textit{bpftime}, steering its development towards addressing the identified challenges while maximizing the utility and performance of eBPF in user space.

\subsection{Overview}

\begin{figure}[t]
\centering
\includegraphics[width=0.5\textwidth]{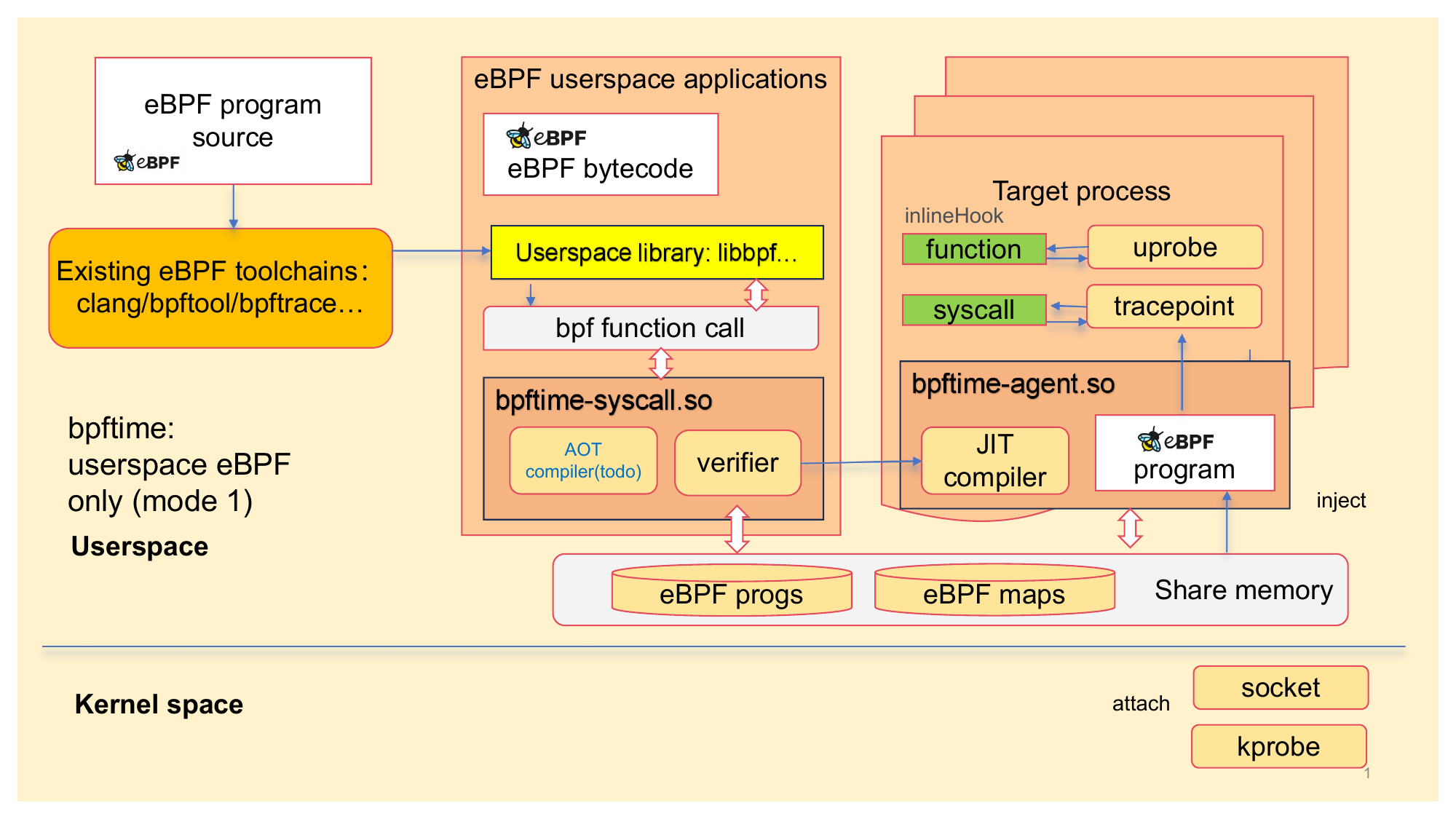}
\caption{The Workflow of kernel eBPF runtime}\label{fig:bpftime}
\end{figure}

The architectural blueprint of the runtime is illustrated in \cref{fig:bpftime}.

The runtime chiefly comprises two components: a syscall compatible library and an attachment agent. The syscall compatible library interfaces with user space eBPF applications, translating eBPF-related system calls into function calls. This library takes the helm of creating shared memory, placing bpf program instructions within it, and initializing hash maps in this shared memory. This design ensures that user space eBPF applications, such as monitoring daemons or tracers typically written using libbpf, can function without the need for recompilation or specifying a particular library version.

Upon the initiation of the eBPF application, and the subsequent creation of the eBPF program and maps in shared memory, the attachment agent swings into action. This agent, embodied as a shared library, can be dynamically injected into the target running process without necessitating recompilation, manual integration of runtime through glue code, or program restarts. The target process is momentarily paused using ptrace to facilitate the loading of the shared library. Once loaded, the shared library establishes attachments to system calls or user space functions within the process's address space. When the program encounters the designated function, control is handed over to the user space eBPF program. This user space eBPF program can then interact with the host environment by updating user space maps, utilizing user space eBPF helpers, or invoking user space Foreign Function Interface (FFI) functions, all without incurring kernel context switches or system calls, thereby enhancing performance and reducing overhead.

In bpftime, the system is designed such that while the host application and the eBPF runtime share the same address space, they do not share libraries, including libc. This ensures that there are distinct instances of libc for each component, which helps to avoid problematic interactions. For instance, if a non-reentrant function from libc is called by the application, and then a code path that invokes the JIT compiler is triggered, the separate libc instances prevent re-entry into the same function in an inconsistent state, averting potential errors. By isolating the libraries in this manner, bpftime maintains separate namespaces for each component, thus enhancing stability and reliability.

The design encapsulated here underscores a meticulous blend of operational efficacy and ease of integration, ensuring the runtime can seamlessly dovetail with existing processes and eBPF applications. The coupling of a syscall-compatible library with a dynamic attachment agent not only simplifies the deployment of user space eBPF applications but also paves the way for efficient runtime interactions sans the performance penalties traditionally associated with kernel context switches and system calls.

\subsection{Injection Design}
\label{subsec:injection_design}

The injector component of bpftime is responsible for loading the eBPF runtime into the address space of a target application. Utilizing the Unix \texttt{ptrace} API, the injector commandeers the application, allowing it to seize the processor context. It then maps the runtime shared library into the application's address space, kick-starting the execution. Upon initialization, bpftime fetches eBPF programs from shared memory, conducts verification, and commences their execution.

Employing \texttt{ptrace} enables the attachment to running processes akin to a debugger, with the added capability to detach and revert to the original code base if required. Post initialization, the control plane application interfaces with the runtime library via shared memory maps or direct kernel mappings, facilitating communication.

An alternative injection method provided by bpftime is the use of the \texttt{LD\_PRELOAD} environment variable. This method coerces the dynamic linker to preload the bpftime shared library into the address space, offering another vector for runtime integration.

\subsection{Hook Design}

In \textit{bpftime}, we amalgamate several dynamic tracing techniques in userspace, predominantly leveraging binary rewriting akin to Just-In-Time (JIT) compilation. This approach facilitates on-the-fly modification of process memory during runtime, obviating the need for recompilation or alteration of the program's source code. However, there are certain constraints to this method. For instance, if the userspace application operates within a JIT-enabled virtual machine, or if prior hooks have modified the target instructions, the hooking process may falter, leading to unforeseen consequences.

\paragraph{Function Hooks}

The design of function hooks employs inline hooking methodology, where the initial instructions at the target function are conserved and substituted with call instructions redirecting to our defined routine. This method ensures the preservation of the original function's behavior while allowing the interception and, potentially, modification of its execution flow. The entry instructions at the target function are replaced with call instructions that divert the execution flow to our hook function. The hook function can then execute additional code, modify the function's parameters, or conditionally alter the execution flow before returning control to the original function.

\paragraph{System Hooks}

System hooking presents a more intricate challenge due to the variance in syscall execution. While syscalls are typically executed within libc, they can also be invoked within other dynamic libraries or the program itself. Addressing this, our approach necessitates a traversal through all code sections in the address space, modifying syscall instructions to redirect them to our eBPF runtime. 

For ARM architecture, syscall hooking mirrors the inline hooking method, enabling direct instruction replacement. However, on x86, the syscall instruction (\textit{sysenter}) spans merely two bytes, which precludes the direct placement of branch or call instructions. To circumvent this limitation, we devised a zepoline method, inspired by previous works\cite{yasukata2023zpoline}, utilizing the zero page to accommodate call instructions within the two-byte constraint. This zepoline method forms a bridge, redirecting the syscall execution to our eBPF runtime, thereby facilitating the desired syscall hooking on x86 platforms.

Through these hook designs, \textit{bpftime} not only accommodates dynamic tracing in userspace but also ensures a robust mechanism for function and system hooking across different architectures, enhancing the versatility and applicability of our userspace eBPF runtime.

\section{Implementation}

We have developed \textit{bpftime} as a prototype of a userspace eBPF runtime capable of executing kernel-compatible real-world eBPF programs, such as bcc-tools and Prometheus eBPF exporter, within a userspace environment for practical workloads. 

The implementation and test framework comprises approximately 8166 LOC of C++, 12733 lines of C, and 992 lines of Python.

\subsection{eBPF VM and JIT Compilation}
Our implementation leverages LLVM JIT (Just-In-Time) for the compilation process. Initially, the eBPF instructions are translated to LLVM Intermediate Representation (IR), which are then compiled to native code by the LLVM backend. Additionally, we have adapted \textit{ubpf} to create a VM (Virtual Machine) with Software Fault Isolation (SFI) options, which can be enabled as needed. For constrained devices, we developed a simplistic handcrafted JIT compiler for x86 architecture to ensure smooth operation.

\subsection{Hook Methods}
The hook methods are crafted with the assistance of Frida\cite{fridagum} and libcapstone. We manually implemented these methods, translating the instructions and patching the target process at runtime to support syscall hooks or function uprobes. A crucial aspect of this implementation is the collection of register states, which are then passed to the eBPF VM. This ensures that uprobes can accurately access the function parameters and return values during execution.

\subsection{Shared Memory Implementation}
The shared memory functionality is realized with the Boost library. We have devised a mechanism to capture the necessary syscalls for loading and attaching eBPF programs, as well as for opening and operating the BPF maps as facilitated by libbpf or other userspace eBPF libraries. The states are preserved in shared memory and replayed when an agent dynamically attaches to the process. For the implementation of syscall hooks related to BPF, two approaches were employed: utilizing \textit{ld preload} in dynamic linking or hooking the syscall as described earlier.

\subsection{Integration and Testing}
The real essence of \textit{bpftime}'s capability is illuminated when integrated with actual workloads. Through meticulous testing, we have ensured that \textit{bpftime} seamlessly interfaces with existing eBPF toolchains and real-world eBPF programs, underscoring its potential to bridge the gap between kernel and userspace eBPF execution while maintaining compatibility and ease of use.

Our implementation endeavors not only to bring forth a novel userspace eBPF runtime but also to establish a solid foundation for further exploration and optimization in this domain, nudging closer to a robust and versatile userspace eBPF ecosystem.

\section{Security Architecture and Threat Mitigation}

\label{sec:security}

This section outlines the comprehensive security measures implemented in bpftime to mitigate a range of potential threats, ensuring the integrity, confidentiality, and availability of the user-space eBPF runtime.

\subsection{Threat Model}
\label{subsec:threat-model}

The threat model for bpftime accounts for scenarios where the host process, potentially compromised or manipulated, may attempt to disrupt the eBPF runtime encapsulated within the shared library. This includes the risk of the host process executing arbitrary code that could interfere with the operation of eBPF programs, leading to memory corruption, privilege escalation, or data leakage. Furthermore, since bpftime allows eBPF runtime to access kernel eBPF maps, there is a tangible risk that the process could inadvertently or maliciously manipulate these maps, affecting the kernel's eBPF operations. Mitigation strategies focus on rigorous access control, process and memory isolation, and continuous integrity checks to ensure that even if the eBPF runtime's immediate environment is compromised, the impact remains contained without spreading to the control plane, other eBPF agents, or the kernel's eBPF subsystem.
 \subsection{Security Mechanisms}
\label{subsec:security-mechanisms}

Bpftime incorporates a multi-layered security architecture to ensure the robustness and integrity of the eBPF runtime. The security properties are defined as follows:

\textbf{SP1: Verifier-Ensured Safety}: eBPF is not a sandbox in its design. it minimizes run-time checks, known the programs before execution with verifier to ensure high performance, we can use either kernel eBPF verifier or userspace verifier to guarantee that eBPF programs do not compromise the target process. With comprehensive checks, including the validation of BPF byte code using kernel eBPF verifier or user-space equivalents, we ensure that: (i) eBPF programs are resilient against malicious input; (ii) they maintain type and memory safety, preventing hardware exceptions; (iii) they can access necessary host process memory or data structures safely through function parameters or helper functions. Through Compile Once - Run Everywhere (CO-RE) technology, eBPF programs are correctly relocated even when the underlying software version or data structure offsets change, ensuring consistent access to the correct memory addresses. The eBPF CO-RE mechanism will relocate eBPF code in the user space library, eg. libbpf, and the relocation can be checked in the verifier. 

\textbf{SP2: Runtime Memory Protection}: Bpftime ensures the eBPF runtime memory is sequestered within a shared memory space, fortified against unauthorized modifications or executions. Memory regions of the shared library are set with read-only permissions at the OS level, coupled with code signing and verification to thwart execution of unauthenticated code. Control flow integrity measures are instituted to further safeguard the shared library and JIT-compiled code.

\textbf{SP3: Segregated Shared Memory Management}: The shared memory within bpftime is stratified into distinct sections for program metadata and map data. Control plane mechanisms provision separate shared memory segments for each eBPF map and program, thereby isolating their operations and preventing any cross-contamination or unauthorized access between them. The agent eBPF runtime can only read the bpf programs and metadata section and cannot modify them or delete any section. The agent eBPF runtime can read or write the map data section, which is shared between different processes.

\textbf{SP4: Unprivileged Kernel eBPF Maps Access}: Bpftime facilitates a secure mechanism for userspace processes to interact with kernel eBPF maps without the need for elevated privileges, which contrasts with the standard kernel uprobe operations. Traditionally, such interactions require elevated privileges, such as CAP\_SYS\_ADMIN or CAP\_SYS\_BPF, posing substantial security risks. To circumvent this, bpftime introduces a method for unprivileged access by leveraging the eBPF File System (BPFFS). Following map creation by the eBPF control plane process, typically privileged, the bpftime-daemon can manage these maps and expose them to unprivileged target processes housing the userspace eBPF runtimes. This involves mounting the BPFFS, conventionally located at /sys/fs/bpf, which may necessitate a relaxation of the default security policies or the use of privileged processes. Access to BPFFS enables the process to retrieve the relevant file descriptors. An alternate solution is to establish the filesystem at a custom location other than /sys/fs/bpf, permitting unprivileged process access while control is maintained through traditional file permission settings. After that, the eBPF runtime embeds in userspace target process can only access the necessary kernel eBPF maps related to userspace programs, cannot create new maps, load new eBPF programs or attach to kernel events. This approach underpins the security model for bpftime's map accessibility.

These security mechanisms are systematically integrated into bpftime, forming a comprehensive defense strategy against a spectrum of threats while preserving the functionality and performance of the eBPF runtime.
\section{Evaluation}

In this section, we aim to rigorously evaluate the efficacy, compatibility, and security of our userspace eBPF runtime, \textit{bpftime}, against its kernel counterpart and other userspace runtimes. We structure our evaluation around the following key questions to offer a comprehensive understanding of \textit{bpftime}'s performance and safety profile:

\begin{enumerate}
    \item \textbf{Performance Comparison:} 
    How does the performance of userspace uprobes and syscall hooks in \textit{bpftime} compare to kernel uprobes and syscall hooks?
    
    \item \textbf{Runtime Efficiency:} 
    How does the performance of LLVM JIT/AOT in \textit{bpftime} stack up against other eBPF userspace runtimes, native code or wasm runtimes?
    
    \item \textbf{Compatibility:} 
    How well does userspace eBPF in \textit{bpftime} integrate with kernel eBPF to handle real-world scenarios?
    
    \item \textbf{Security Assessment:}
    How does userspace eBPF in \textit{bpftime} mitigate potential security threats that may arise from kernel eBPF?
\end{enumerate}
\subsection{Hook Performance Comparison}

We conducted micro-benchmarks to compare the performance of userspace and kernel uprobes. The results are summarized in the table below, indicating the latency (in nanoseconds) and instruction count for each probe/tracepoint type.

\begin{table}[H]
    \centering
    \begin{tabular}{|c|c|c|c|}
        \hline
        Probe Types & Kernel (ns) & User (ns) & \#Inst \\
        \hline
        Uprobe & 3224.172760 & 314.569110 & 4 \\
        Uretprobe & 3996.799580 & 381.270270 & 2 \\
        Syscall Tracepoint & 151.82801 & 232.57691 & 4 \\
        Embedding Runtime & Not available &  110.008430 & 4 \\
        \hline
    \end{tabular}
    \caption{Performance comparison of kernel and userspace probes.}
    \label{tab:probe_comparison}
\end{table}

The userspace uprobes and uretprobes in \textit{bpftime} show a substantial reduction in latency compared to the kernel counterparts. Specifically, the userspace uprobe latency is about 314.57 ns, while the kernel uprobe latency is showcasing a 10x improvement. Similarly, the userspace uretprobe performs better with a latency nearly 10x compared to the kernel uretprobe latency.

However, for syscall tracepoints, the kernel implementation is more efficient with a latency of 151.83 ns, as opposed to the userspace implementation with a latency of 232.58 ns, which is acceptable.

The 'Embedding runtime' metric demonstrates the latency of embedding the userspace eBPF runtime within a process, showing a promising latency of 110.01 ns. 

The instruction count indicates the complexity of each operation, which remains consistent across both kernel and userspace implementations for the respective probe and tracepoint types.

These findings indicate that \textit{bpftime} offers a significant performance advantage for uprobes and uretprobes, while maintaining a comparable level of complexity, making it a compelling alternative for kernel uprobes and uretprobes.

\subsection{Runtime Efficiency}

We assess the efficiency of LLVM JIT/AOT in \textit{bpftime} by comparing it with other eBPF userspace runtimes, native code, and wasm runtimes using a set of micro-benchmarks. These benchmarks include integer computations (`log2\_int`), prime number generation (`prime`), memory copying (`memcpy`), simple control flow (`simple`), switch statement (`switch`), string copy that fails (`strcmp\_fail`), memory accessing (`memory\_a\_plus\_b`).

\begin{figure*}[t]
\centering
\includegraphics[width=\textwidth]{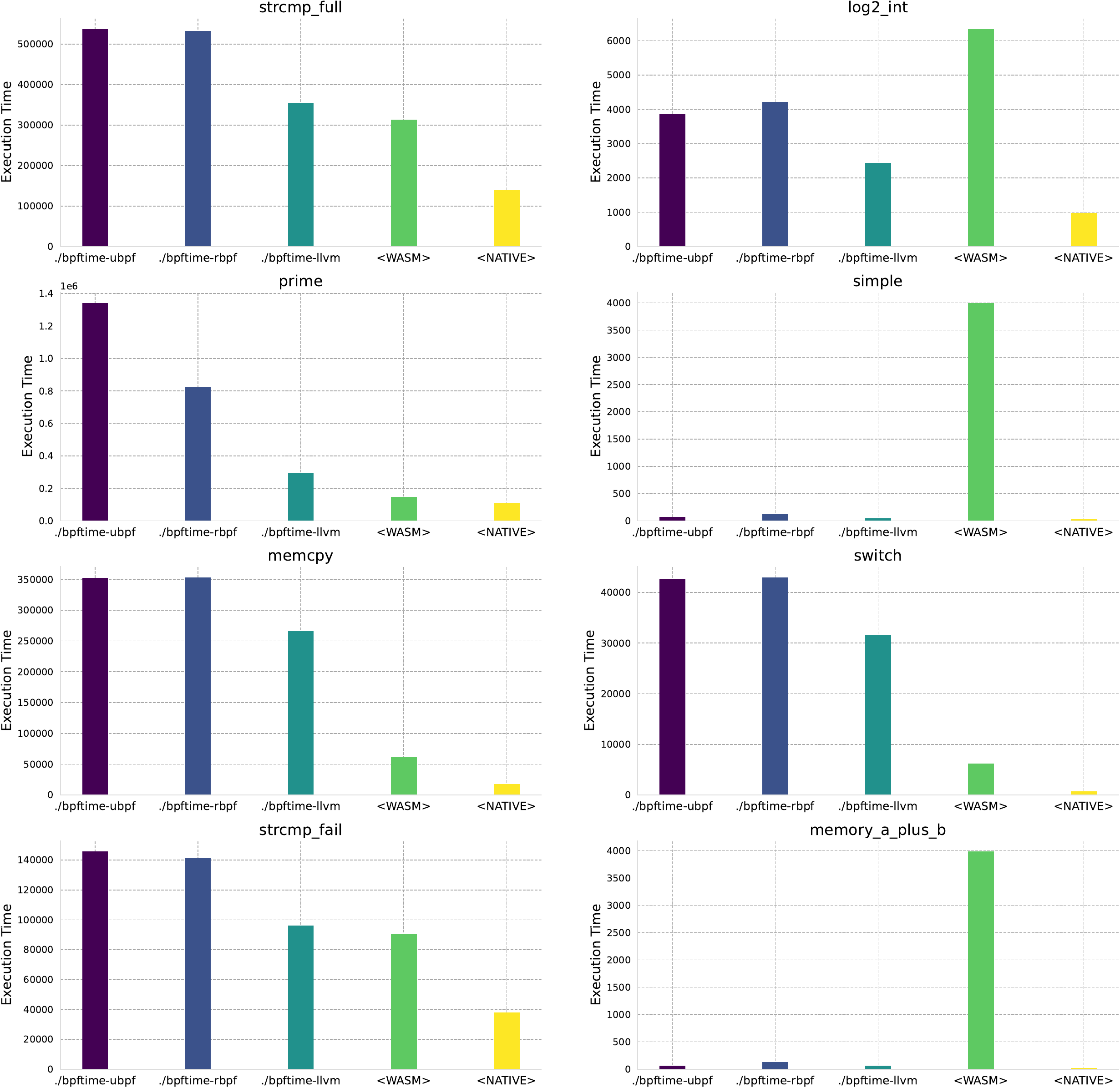}
\caption{Performance comparison of LLVM JIT in \textit{bpftime} with other runtimes}\label{fig:bench-llvm-jit}
\end{figure*}

Figure \ref{fig:bench-llvm-jit} demonstrates that LLVM JIT/AOT in \textit{bpftime} outperforms other runtimes across all tested scenarios, particularly excelling in integer computations and complex mathematical operations. The results also indicate optimized memory handling in memory operations by \textit{bpftime}. 

The superior performance of LLVM JIT in \textit{bpftime} not only highlights its computational efficiency but also its execution speed, making it a preferable choice for applications demanding rapid execution and high computational throughput. This analysis aids users in selecting the most suitable runtime for their specific use-cases, underscoring the efficiency of \textit{bpftime} as a viable userspace eBPF runtime.
\subsection{Compatibility}

We conduct a compatibility analysis of \textit{bpftime} by deploying a set of real-world eBPF programs from bcc, targeting both userspace and kernel space probing. The evaluation emphasizes ease of migration, feature parity, and seamless integration in practical applications.

Utilizing \textit{bpftime} allows for:
\begin{itemize}
    \item \textbf{Userspace Function Tracing:}
    By attaching uprobe, uretprobe, or syscall tracepoints (currently supported on x86), eBPF programs can be directed towards a process or a collection of processes. For instance, the \texttt{malloc} example demonstrates tracing \texttt{malloc} calls within libc by process ID, showcasing basic \texttt{hashmap} utilization with userspace \texttt{uprobe}.
    
    \item \textbf{Syscall Tracing:}
    By utilizing syscall tracepoints, file open or close syscalls within a process can be traced as seen in the \texttt{opensnoop} example. This showcases the use of userspace \texttt{syscall tracepoint} alongside \texttt{ring buffer} output for data collection.
\end{itemize}

The examined eBPF programs run unmodified on \textit{bpftime}, indicating a high level of compatibility and ease of transition from kernel eBPF runtime to \textit{bpftime}. This seamless migration, coupled with the functional equivalence demonstrated in the micro-benchmarks, posits \textit{bpftime} as a viable userspace eBPF runtime for real-world use cases.
\subsection{Security Assessment}

We analyze the security posture of userspace eBPF in \textit{bpftime} by exploring the potential attack vectors and the measures \textit{bpftime} employs to thwart such threats when compared to kernel eBPF.

\section{Related Work}

\subsection{Dynamic Binary Instrumentation}

Dynamic Binary Instrumentation (DBI) facilitates real-time analysis of binary applications by injecting instrumentation code during runtime. This mechanism permits the embedding of specific analysis code into a program as it executes, tailored to user specifications, without disrupting the dynamic execution flow of the program. Renowned platforms that provide dynamic binary analysis capabilities include Pin\cite{1550984}, DynamoRIO\cite{dynamorio}, and Frida\cite{fridagum}. The defining attributes of DBI encompass its capability to work directly on binary code, obviating the need for source code, and its dynamic modus operandi which allows on-the-fly modifications to the program, thus negating the necessity for recompilation or program restarts. Nonetheless, DBI comes with certain caveats: it lacks built-in security checks or verification measures which may pose potential security risks. Moreover, in the absence of a high-performance virtual machine (VM) and an encapsulated sandbox environment, the efficiency of DBI could be compromised, and there's a potential to disrupt the execution states of the target process. Furthermore, existing DBI tools fall short in providing a high-performance control panel or data aggregation mechanisms akin to the capabilities offered by eBPF maps.

\subsection{WebAssembly}

WebAssembly (Wasm)\cite{webassembly} is a binary instruction format created as a portable compilation target for high-level languages such as C, C++, and Rust. It endeavors to achieve execution at native speed by leveraging common hardware capabilities. WebAssembly's design ensures portability across all modern platforms, both on desktop and mobile. It also operates within a sandboxed execution environment to maintain secure execution. With its binary format, WebAssembly enables faster parsing, compiling, and execution compared to traditional text-based web technologies. The structure and principles behind WebAssembly make it a suitable platform for various use-cases beyond the web, including blockchain, serverless computing, and portable CLI tools, among others.

Both Dynamic Binary Instrumentation and WebAssembly illustrate unique approaches to runtime code analysis and execution. While DBI offers detailed insights into binary code behavior during runtime, WebAssembly provides a secure and portable compilation target. The exploration of these technologies, alongside the development of \textit{bpftime}, contributes to the broader effort of advancing runtime code analysis and execution frameworks.
\section{Conclusion}

In this work, we presented \textit{bpftime}, a cutting-edge userspace eBPF runtime designed to alleviate the identified performance and security drawbacks inherent to kernel eBPF. Through \textit{bpftime}, we significantly improved Uprobe and Syscall hook performance, demonstrating a tenfold reduction in Uprobe overhead relative to kernel uprobes. Our runtime showcases a smooth integration with ongoing processes, bypassing the need for restarts or manual recompilations. Compatibility with prevalent eBPF toolchains like clang and libbpf simplifies userspace eBPF development, portraying \textit{bpftime} as a step forward in enhancing userspace eBPF runtimes. The open-sourced nature of \textit{bpftime} at \href{https://github.com/eunomia-bpf/bpftime}{https://github.com/eunomia-bpf/bpftime} paves the way for community engagement, fostering further exploration and innovation in this domain.

\bibliographystyle{abbrvnat}

\bibliography{cite}

\begin{thebibliography}{20}
\providecommand{\natexlab}[1]{#1}
\providecommand{\url}[1]{\texttt{#1}}
\expandafter\ifx\csname urlstyle\endcsname\relax
  \providecommand{\doi}[1]{doi: #1}\else
  \providecommand{\doi}{doi: \begingroup \urlstyle{rm}\Url}\fi

\bibitem[Authors()]{webassembly}
W.~Authors.
\newblock Webassembly specifications.
\newblock \url{https://webassembly.github.io/spec/}.

\bibitem[DynamoRIO()]{dynamorio}
DynamoRIO.
\newblock Dynamic instrumentation tool platform.
\newblock \url{https://github.com/DynamoRIO/dynamorio}.

\bibitem[frida()]{fridagum}
frida.
\newblock Cross-platform instrumentation and introspection library written in
  c.
\newblock \url{https://github.com/frida/frida-gum}.

\bibitem[gojue()]{ecapture}
gojue.
\newblock Capture ssl/tls text content without a ca certificate using ebpf.
\newblock \url{https://github.com/gojue/ecapture}.

\bibitem[He et~al.(2022{\natexlab{a}})He, Zou, Sun, Liu, Xu, Wang, Shen, Wang,
  and Li]{he2022rapidpatch}
Y.~He, Z.~Zou, K.~Sun, Z.~Liu, K.~Xu, Q.~Wang, C.~Shen, Z.~Wang, and Q.~Li.
\newblock $\{$RapidPatch$\}$: Firmware hotpatching for $\{$Real-Time$\}$
  embedded devices.
\newblock In \emph{31st USENIX Security Symposium (USENIX Security 22)}, pages
  2225--2242, 2022{\natexlab{a}}.

\bibitem[He et~al.(2022{\natexlab{b}})He, Zou, Sun, Liu, Xu, Wang, Shen, Wang,
  and Li]{rapidpatch}
Y.~He, Z.~Zou, K.~Sun, Z.~Liu, K.~Xu, Q.~Wang, C.~Shen, Z.~Wang, and Q.~Li.
\newblock {RapidPatch}: Firmware hotpatching for {Real-Time} embedded devices.
\newblock In \emph{31st USENIX Security Symposium (USENIX Security 22)}, pages
  2225--2242, Boston, MA, Aug. 2022{\natexlab{b}}. USENIX Association.
\newblock ISBN 978-1-939133-31-1.
\newblock URL
  \url{https://www.usenix.org/conference/usenixsecurity22/presentation/he-yi}.

\bibitem[He et~al.(2023)He, Guo, Xing, Che, Sun, Liu, Xu, and Li]{he2023cross}
Y.~He, R.~Guo, Y.~Xing, X.~Che, K.~Sun, Z.~Liu, K.~Xu, and Q.~Li.
\newblock Cross container attacks: The bewildered $\{$eBPF$\}$ on clouds.
\newblock In \emph{32nd USENIX Security Symposium (USENIX Security 23)}, pages
  5971--5988, 2023.

\bibitem[iovisor({\natexlab{a}})]{dpdk-ebpf}
iovisor.
\newblock Userspace ebpf vm, {\natexlab{a}}.
\newblock \url{https://doc.dpdk.org/guides/prog_guide/bpf_lib.html}.

\bibitem[iovisor({\natexlab{b}})]{ubpf}
iovisor.
\newblock Userspace ebpf vm, {\natexlab{b}}.
\newblock \url{https://github.com/iovisor/ubpf}.

\bibitem[Keniston et~al.(2007)Keniston, Mavinakayanahalli, Panchamukhi, and
  Prasad]{keniston2007ptrace}
J.~Keniston, A.~Mavinakayanahalli, P.~Panchamukhi, and V.~Prasad.
\newblock Ptrace, utrace, uprobes: Lightweight, dynamic tracing of user apps.
\newblock In \emph{Proceedings of the 2007 Linux symposium}, pages 215--224,
  2007.

\bibitem[Lim et~al.(2023)Lim, Han, and Pasquier]{lim2023unleashing}
S.~Y. Lim, X.~Han, and T.~Pasquier.
\newblock Unleashing unprivileged ebpf potential with dynamic sandboxing.
\newblock In \emph{Proceedings of the 1st Workshop on eBPF and Kernel
  Extensions}, pages 42--48, 2023.

\bibitem[microsoft()]{windows-ebpf}
microsoft.
\newblock ebpf for windows.
\newblock \url{https://github.com/microsoft/ebpf-for-windows}.

\bibitem[Patil et~al.(2004)Patil, Cohn, Charney, Kapoor, Sun, and
  Karunanidhi]{1550984}
H.~Patil, R.~Cohn, M.~Charney, R.~Kapoor, A.~Sun, and A.~Karunanidhi.
\newblock Pinpointing representative portions of large intel ® itanium ®
  programs with dynamic instrumentation.
\newblock In \emph{37th International Symposium on Microarchitecture
  (MICRO-37'04)}, pages 81--92, 2004.
\newblock \doi{10.1109/MICRO.2004.28}.

\bibitem[Project(2023)]{bcc}
I.~V. Project.
\newblock Bpf compiler collection (bcc), 2023.
\newblock Available: \url{https://github.com/iovisor/bcc}.

\bibitem[qmonnet()]{rbpf}
qmonnet.
\newblock Rust virtual machine and jit compiler for ebpf programs.
\newblock \url{https://github.com/qmonnet/rbpf}.

\bibitem[Shen et~al.(2023)Shen, Zhang, Xiang, Shi, Li, Shen, Zhang, Wu, Yin,
  Wang, et~al.]{shen2023network}
J.~Shen, H.~Zhang, Y.~Xiang, X.~Shi, X.~Li, Y.~Shen, Z.~Zhang, Y.~Wu, X.~Yin,
  J.~Wang, et~al.
\newblock Network-centric distributed tracing with deepflow: Troubleshooting
  your microservices in zero code.
\newblock In \emph{Proceedings of the ACM SIGCOMM 2023 Conference}, pages
  420--437, 2023.

\bibitem[Yasukata et~al.(2023)Yasukata, Tazaki, Aublin, and
  Ishiguro]{yasukata2023zpoline}
K.~Yasukata, H.~Tazaki, P.-L. Aublin, and K.~Ishiguro.
\newblock zpoline: a system call hook mechanism based on binary rewriting.
\newblock In \emph{2023 USENIX Annual Technical Conference (USENIX ATC 23)},
  pages 293--300, 2023.

\bibitem[Zandberg and Baccelli(2021)]{femto}
K.~Zandberg and E.~Baccelli.
\newblock Femto-containers: Devops on microcontrollers with lightweight
  virtualization \& isolation for iot software modules.
\newblock \emph{arXiv preprint arXiv:2106.12553}, 2021.

\bibitem[Zandberg et~al.(2022)Zandberg, Baccelli, Yuan, Besson, and
  Talpin]{zandberg2022femto}
K.~Zandberg, E.~Baccelli, S.~Yuan, F.~Besson, and J.-P. Talpin.
\newblock Femto-containers: Lightweight virtualization and fault isolation for
  small software functions on low-power iot microcontrollers.
\newblock In \emph{Proceedings of the 23rd ACM/IFIP International Middleware
  Conference}, pages 161--173, 2022.

\bibitem[Zhong et~al.(2022)Zhong, Li, Wu, Zarkadas, Tao, Mesterhazy, Makris,
  Yang, Tai, Stutsman, and Cidon]{Zhong22}
Y.~Zhong, H.~Li, Y.~J. Wu, I.~Zarkadas, J.~Tao, E.~Mesterhazy, M.~Makris,
  J.~Yang, A.~Tai, R.~Stutsman, and A.~Cidon.
\newblock {XRP}: {In-Kernel} storage functions with {eBPF}.
\newblock In \emph{16th USENIX Symposium on Operating Systems Design and
  Implementation (OSDI 22)}, pages 375--393, Carlsbad, CA, July 2022. USENIX
  Association.
\newblock ISBN 978-1-939133-28-1.
\newblock URL
  \url{https://www.usenix.org/conference/osdi22/presentation/zhong}.

\end{thebibliography}

\end{document}